\documentclass[prd,a4paper,superscriptaddress,nofootinbib,twocolumn]{revtex4}
 \usepackage{amsmath,amssymb,epsfig,natbib}

\begin{document}

\title{Observational constraints on the curvaton model of inflation}

\author{Christopher Gordon}
 \email{C.Gordon@damtp.cam.ac.uk}
 \affiliation{DAMTP, CMS, Wilberforce Road, Cambridge CB3 0WA, UK.}
 
\author{Antony Lewis}
 \email{Antony@AntonyLewis.com}
 \affiliation{CITA, 60 St. George St, Toronto M5S 3H8, ON, Canada.}

\newcommand\fig[1]{Fig.~\ref{#1}}
\newcommand{\infl}{\psi}
\newcommand{\curv}{\sigma}
\newcommand\web[1]{#1}

\newcommand{\Hunit}{~\text{km}~\text{s}^{-1} \Mpc^{-1}}

\newcommand{\Mp}{M_{\rm p}}
\newcommand{\zetarad}{\zeta^{\text{rad}}}
\newcommand{\Rb}{R_b}
\newcommand{\Rc}{R_c}
\newcommand{\Rg}{R_\gamma}
\newcommand{\Rv}{R_\nu}
\newcommand{\Drho}{\hat{\Delta}}
\newcommand\ba{\begin{eqnarray}}
\newcommand\ea{\end{eqnarray}}
\newcommand\be{\begin{equation}}
\newcommand\ee{\end{equation}}
\newcommand\lagrange{{\cal L}}
\newcommand\cll{{\cal L}}
\newcommand\clx{{\delta\rho}}
\newcommand\clz{{\cal Z}}
\newcommand\clv{{\cal V}}
\newcommand\clo{{\cal O}}
\newcommand\cla{{\cal A}}
\newcommand\clr{{\cal R}}
\newcommand\clp{{\cal P}}

\newcommand\del{\nabla}
\newcommand\Tr{{\rm Tr}}
\newcommand\half{{\frac{1}{2}}}
\renewcommand\H{{\cal H}}
\newcommand\K{{\cal K}}

\renewcommand\P{{\cal P}}
\newcommand{\Mpc}{\text{Mpc}}

\newcommand{\la}{\langle}
\newcommand{\ra}{\rangle}
\newcommand{\Omtot}{\Omega_{\mathrm{tot}}}

\newcommand{\R}{{\mbox{$\chi$}}}
\renewcommand{\S}{\mbox{$\cal S$}}
\newcommand{\Sc}{\mbox{${\cal S}_c$}}
\newcommand{\Sb}{\mbox{${\cal S}_b$}}
\newcommand{\Snu}{\mbox{$\cal S_{\nu}$}}
\newcommand{\Sbeff}{\S^{\text{eff}}_{{b}}}

\newcommand{\Omegab}{\mbox{$\Omega_{b}$}}
\newcommand{\Omegac}{\mbox{$\Omega_{c}$}}
\newcommand{\Rprim}{\mbox{${\cal R}_{prim}$}}
\newcommand{\etal}{et.~al.\ }

\newcommand\eq[1]{Eq.~(\ref{#1})}
\newcommand\eqs[2]{Eqs.~(\ref{#1}) and (\ref{#2})}
\newcommand\eqss[3]{Eqs.~(\ref{#1}), (\ref{#2}) and (\ref{#3})}
\newcommand\eqsss[4]{Eqs.~(\ref{#1}), (\ref{#2}), (\ref{#3})
and (\ref{#4})}
\newcommand\eqssss[5]{Eqs.~(\ref{#1}), (\ref{#2}), (\ref{#3}),
(\ref{#4}) and (\ref{#5})}
\newcommand\eqsssss[6]{Eqs.~(\ref{#1}), (\ref{#2}), (\ref{#3}),
(\ref{#4}), (\ref{#5}) and (\ref{#6})}
\newcommand\eqst[2]{Eqs.~(\ref{#1})--(\ref{#2})}
\newcommand\<{\left<}
\renewcommand\>{\right>}

\newcommand\slabel[1]{\label{#1}}

\begin{abstract}
\vspace{\baselineskip}
Simple curvaton models can generate a mixture of 
of correlated primordial adiabatic and isocurvature perturbations. The
baryon and cold dark matter isocurvature modes differ only by an
observationally null mode in which the two perturbations almost exactly
compensate, and therefore have proportional effects at linear order.  
We discuss the CMB anisotropy in general mixed models, and give a
simple approximate analytic result for the
large scale CMB anisotropy.  
Working numerically we use
the latest WMAP observations and
a variety of other data to
constrain the curvaton model.
We find that models with an isocurvature contribution are not favored
relative to simple purely adiabatic models. However a 
significant
primordial totally
correlated baryon isocurvature perturbation 
is not ruled out.  
Certain classes of curvaton model are thereby ruled out,
other classes predict enough
non-Gaussianity to be detectable by the Planck satellite.  In the
appendices we review the relevant equations in the covariant
formulation and give series solutions for the radiation dominated era.
\end{abstract}

\maketitle


\section{Introduction}
Recent detailed measurements of the acoustic peaks in CMB anisotropy
power spectrum by the WMAP satellite~\cite{Hinshaw03,Peiris03} 
are consistent with the standard model of a predominantly {\em adiabatic},
approximately scale invariant primordial power spectrum in a spatially flat
Universe. Frequently it is assumed the initial power spectrum is
entirely adiabatic, though there is still no compelling justification
for this assumption. Although adiabatic perturbations are predicted from single field 
models of inflation \cite{LL}, if one allows the possibility of
multiple fields in the early Universe then there is also the possibility
of \emph{isocurvature} perturbations  (also known as \emph{entropy}
perturbations)
\cite{Kofman87,Polarski94,Garcia-Bellido96,Linde97,Sasaki98,Langlois99,Gordon00,Hwang00,Starobinsky01,Bartolo01,GrootNibbelink01,Wands02,Tsujikawa02}.
In particular, the recently
proposed {\em curvaton} model uses a second scalar field (the
`curvaton')  to form the
perturbations \cite{Mollerach90,Lyth01,Lyth02,Moroi01,Moroi02, Murayam02}.
The motivation for this is it makes it easier for otherwise
satisfactory particle physics models of inflation to produce the
correct primordial spectrum of perturbations
\cite{Dimopoulos01}. 
Various candidates for the curvaton have been 
proposed \cite{Postma02,Enqvist02,Bastero-Gil02,Dimopoulos03a,Dimopoulos03b}. 
A curvaton mechanism has also been considered in the pre big bang
scenario \cite{Copeland97,Lidsey00,Enqvist01,Sloth02,Bozza02}
where it can be used to produce an almost scale invariant spectrum.

The curvaton scenario also has the feature of being able to
generate isocurvature perturbations of a similar magnitude to the
adiabatic perturbation without fine tuning, and therefore is open to 
observational test.

Early studies of non-adiabatic perturbations, either 
considered 
purely isocurvature cold dark matter perturbations \cite{Efstathiou86} or
mixtures of adiabatic and uncorrelated cold dark matter isocurvature
perturbations \cite{Stompor96,Pierpaoli99,Kawasaki2001,Enqvist2000}. 
However, as first realized by
Langlois \cite{Langlois99}, the adiabatic and isocurvature components
can be correlated and this correlation may have interesting
observational consequences \cite{Langlois00}. 
In Ref.~\cite{Bucher99} they
identified four regular isocurvature modes, which in general can have
arbitrary correlations with each other and with the adiabatic mode. 
Such general models have many degeneracies
and are badly constrained by pre-WMAP data \cite{Trotta01,Trotta02}.
Detailed CMB polarization data is expected to help with this
\cite{Bucher01}.
In Ref.~\cite{Peiris03} (following Ref.~\cite{Amendola01} with pre-WMAP
data) they 
considered a CDM isocurvature mode
with an arbitrary correlation to an adiabatic mode
and found that though not favored by the data, a significant isocurvature contribution was still permitted.
Constraints on a specific model that
doesn't produce isocurvature modes were given in Ref.~\cite{Bartolo02}.

Here we start in Section~\ref{CMB} by making some general remarks about mixed isocurvature models,
and discuss the corresponding CMB power spectra predictions.  
Then in Section~\ref{curvaton} we discuss
current observational constraints on totally
correlated (or anti-correlated) adiabatic and isocurvature
perturbations, as predicted by the curvaton model. Various scenarios 
within the curvaton model predict specific ratios of adiabatic and
isocurvature perturbations, and can be tested directly. In general we find
constraints on when the curvaton decayed. 

We use the CMB temperature and temperature-polarization cross-correlation
anisotropy power spectra from
the
WMAP\footnote{\url{http://lambda.gsfc.nasa.gov/}}~\cite{Verde03,Hinshaw03,Kogut03}
observations, as well as seven almost independent temperature band
powers from ACBAR\footnote{\url{http://cosmologist.info/ACBAR}}~\cite{Kuo02} on smaller scales. In addition we use
data from
the 2dF galaxy redshift survey~\cite{Percival02}, HST
Key Project~\cite{Freedman01}, and
nucleosynthesis~\cite{Burles01} 
using a slightly modified version of the CosmoMC\footnote{\url{http://cosmologist.info/cosmomc}}
Markov-Chain Monte Carlo program, 
as described in Ref.~\cite{cosmomc}.

For simplicity we assume a flat universe with a cosmological
constant, 
uninteracting cold dark matter, and massless neutrinos
evolving according to general relativity.

\section{Primordial perturbations and the CMB anisotropy}
\label{CMB}
It is well known that the 
curvature perturbation,
in the constant density 
or comoving frame (gauge\footnote{In the context of this article the term frame and gauge
are effectively interchangeable. See Appendix~\ref{cov_eqs} for further discussion.
})
is conserved on super-Hubble scales for adiabatic perturbations
\cite{Bardeen80,BST,Lyth85,Kodama84,Mukhanov92}. This is not the case
in the presence of isocurvature modes since these source changes to
the curvature perturbation. However, as shown in Ref.~\cite{Wands00}
(and reviewed in Appendix~\ref{cov_eqs}), in the presence of
isocurvature modes the large scale evolution can still be analysed
easily using the curvature perturbation in the frame 
in which
the density is unperturbed, $\zeta$.  This can be expressed in terms
of the curvature perturbations in the frames in which individual
species are unperturbed $\zeta_i$ using
\begin{equation}
\zeta = \frac{\sum_i \rho_i' \zeta_i}{\sum_i \rho_i'},
\slabel{zetasum}
\end{equation}
where the dash denotes the derivative with respect to conformal time. 
  For non-interacting conserved particle species 
 the individual $\zeta_i$ are conserved on large scales if there is a
 definite equation of state $p_i=p_i(\rho_i)$. In this case the
  evolution of $\zeta$ follows straightforwardly from \eq{zetasum}
  depending on the evolution of the background energy densities. 
An {\em adiabatic} perturbation is one in
which $\zeta_i = \zeta$ for all $i$, in which case $\zeta$ is constant in time on large
scales. The isocurvature perturbations are defined as~\cite{Malik02}
\begin{equation}
\S_{i,j} \equiv 3(\zeta_i - \zeta_j) =  -3\H \left(\frac{\clx_i}{\rho_i'} -
\frac{\clx_j}{\rho_j'} \right)
\slabel{isocurvature}
\end{equation}
where $\clx_i = \rho_i\Delta_i$ (no sum) is the density perturbation in any
frame and $\H$ is the conformal Hubble rate. 
We
 consider a fluid consisting of photons ($\gamma$), massless neutrinos
 ($\nu$), cold baryons ($b$) and cold dark matter (CDM, $c$), where
it is conventional to describe the perturbations with
$j=\gamma$, in which case the second index can be omitted so
$\Sb\equiv \S_{b,\gamma}$, etc. 
The isocurvature perturbations are conserved on large scales where
the photon-baryon coupling is unimportant. The $\zeta_i$ are related
to the fractional density perturbations in the unperturbed curvature
frame 
by $\Drho_i = 3(1+p_i/\rho_i)\zeta_i$, and for matter with
constant equation of state $\Drho_i$ are also conserved on
large scales.  

In general isocurvature perturbations give rise to perturbations in
the density, and the universe is no longer exactly FRW. One exception
to this is when the two matter perturbations exactly compensate, so
$\clx_c = -\clx_b$, in which case the total matter density is
unperturbed, and hence the universe evolves as though there were no
perturbations.
In such a universe the CMB anisotropy would be
dominated by tiny small scale linear effects due to non-zero pressure
of the baryons or dark mater,
and second order effects due to the perturbation in the
electron number density associated with the baryons.
At linear order 
$\clx_c = -\clx_b$, $\zeta=\clx_\gamma =\clx_\nu=0$ is a time
  independent solution to the 
pressureless 
perturbation equations, and adding this solution to any other solution
will make no difference to the linear CMB anisotropy or matter power
spectrum. 
It follows that an initial isocurvature perturbation with $\Sb =1,
\Sc=0$ is observationally 
essentially
indistinguishable from one with $\Sb = 0, \Sc = \rho_b/\rho_c$.

We now derive an approximate analytic form for the large scale CMB
temperature anisotropy in
the presence of primordial isocurvature and adiabatic perturbations.  
Neglecting a local monopole and dipole contribution, taking
recombination to be instantaneous at 
at conformal time
$\tau_*$, and assuming the reionization
optical depth is negligible,  
the monopole and Integrated Sachs Wolfe (ISW) contributions to the temperature anisotropies due to scalar perturbations are given by
\begin{equation}
\frac{\delta T}{T} \approx\left[ \frac{1}{4} \Drho_\gamma +
  2\phi\right]_{\tau_*} + 2 \int_{\tau_*}^{\tau_0} \phi'\,d\lambda,
\label{deltat}
\end{equation}
where $\phi$
is the Weyl potential (see Appendix~\ref{cov_eqs}) and
 $\tau_0$ is the conformal time today and the integral is along
the comoving photon line of site. 
Additional terms which arise due to the velocities, quadrupoles and
polarization at last scattering are generally sub-dominant on large scales.
Since the pressures are assumed to be zero the perturbations
are purely adiabatic in the matter era, and hence $\zeta$ is constant
on large scales.  During matter domination the potential evolves as
$\phi = C_1 + C_2/\tau^5$, and can be related to $\zeta$ using
\eq{chi_eq} when the anisotropic stress is negligible
\begin{equation}
\phi = -\frac{3}{5}\zeta + \frac{C_2}{\tau^5}.
\end{equation}
Since the radiation to matter density ratio only falls off as $\sim
1/\tau^2$ when the matter dominates, when the approximation of matter
domination is accurate it should also be valid to neglect the decaying
mode $\propto 1/\tau^5$ and assume $\phi \approx -\frac{3}{5}\zeta$.
Using $\Drho_\gamma  = 4\zeta_\gamma$, and neglecting the ISW
contribution \eq{deltat} then  becomes
\begin{equation}
\frac{\delta T}{T} \approx \zeta_{\gamma} - \frac{6}{5} \zeta
\slabel{SW1}
\end{equation}
where from \eq{zetasum} in matter domination
\begin{equation}
\zeta \approx \Rb \zeta_b + \Rc \zeta_c. 
\slabel{zetamat}
\end{equation}
Here we define the matter fractions $\Rb\equiv\rho_b/\rho_m$, $\Rc
\equiv \rho_c/\rho_m$, where $\rho_m = \rho_c + \rho_b$.

During early radiation domination $\zetarad
\approx \Rg \zeta_\gamma + \Rv\zeta_\nu$ from \eq{zetasum}, where we
define the radiation density fractions $\Rv\equiv
\rho_\nu/(\rho_\gamma+\rho_\nu), \Rg \equiv
\rho_\gamma/(\rho_\gamma+\rho_\nu)$. Using \eqs{isocurvature}{SW1} and
the constancy of the large scale $\zeta_i$ we
can therefore relate the large scale temperature anisotropy to the primordial
adiabatic and isocurvature perturbations
\begin{equation}
\frac{\delta T}{T} \approx -\frac{1}{5}\zetarad
-\frac{2}{5} \left( 
\Rc \Sc+ \Rb \Sb    
\right) + \frac{1}{15} \Rv\Snu.
\slabel{SW2}
\end{equation}
We can take $\zetarad$ as a measure of the primordial adiabatic
perturbation. So this formula shows the effect of a mixture of
adiabatic and isocurvature perturbations on the observed large scale
CMB temperature anisotropy.
This result agrees with that in \cite{Langlois00}, 
despite errors in their derivation which arise from an invalid ansatz
for the time evolution of the velocities and anisotropic
stress (demonstrated by counterexample in Appendix~\ref{series}). However
in matter domination the velocities are negligible so the error is
harmless, and for the adiabatic and neutrino modes the assumption is correct to the
required order during radiation domination. However unfortunately their general
result for the evolution of the potential is incorrect and cannot 
be used to improve on the above much simpler
argument.

This analytic argument shows 
the main qualitative features,
though in reality recombination is far from being completely matter
dominated, and the ISW and other contributions will not be
negligible. It is however straightforward to compute the CMB and
matter power spectra numerically \cite{Seljak96,Lewis99} starting from a series solution
in the early radiation dominated era (Appendix~\ref{series}).

\section{Constraining the Curvaton model}
\label{curvaton}
The curvaton scenario provides a mechanism for allowing the 
inflation potential to have more natural properties, at the expense of introducing
an additional unidentified scalar field which generates the perturbations.
In the curvaton model the inflaton field drives the initial expansion and
generates an era of radiation domination after it decays. 
The expansion rate then slows and the curvaton field can
reach the minimum of its potential and start to oscillate. During
oscillation the
curvaton field acts effectively like a matter component, and its   
perturbation acts like a matter isocurvature mode. As the radiation
redshifts further the equation of state then changes to matter domination as
the curvaton density comes to dominate. As the background equation of
state changes, a
curvature perturbation is generated from the isocurvature mode.  The
curvaton then decays into (predominantly) radiation well before nucleosynthesis, and we
enter the usual primordial radiation dominated epoch. 

Primordial
correlated isocurvature modes can be generated 
if the baryons or CDM are generated by, or before, the curvaton
decays, as discussed in detail below.
If one or both were created before the curvaton decays, the
current model assumes that the curvaton had a negligible density when
they decayed \cite{Lyth02}. We assume that the curvaton is the only cosmologically relevant
scalar field after inflaton decay, and that the perturbations in the inflaton field are negligible.
Generically such models predict a very small tensor mode contribution,
which we assume can be neglected. 
We assume there is no lepton number at neutrino decoupling so that there
are no neutrino isocurvature modes, though see~\cite{Lyth02} for other possibilities.

As discussed in Section~\ref{CMB}, the baryon and CDM
isocurvature modes predict proportional results, so we can 
account for $\Sc$ by using just an effective
baryon isocurvature perturbation
\begin{equation}
\Sbeff = \Sb + \frac{\Rc}{\Rb} \Sc.
\slabel{bcdm}
\end{equation}

The baryon and CDM isocurvature perturbations are completely
correlated (or anti-correlated) with each other and the adiabatic
perturbation, so $\Sbeff = B\zetarad$ where $B$ measures the 
isocurvature mode contribution and is taken to be scale
independent\footnote{
Our sign convention for $B$ differs from that in
Ref.~\cite{Amendola01}. In our convention $B>0$ corresponds to a
positive correlation and the modes contribute with the same sign to
the large scale CMB anisotropy.
}.
From \eq{SW2}, the large scale CMB anisotropy variance is then given approximately by
\begin{equation}
\left\la \frac{\delta T^2}{T^2} \right\ra 
\approx \frac{1}{25} (1+2\Rb B)^2 \P_\zeta,
\slabel{SWsq}
\end{equation}
where $\P_\zeta=\P_\zetarad$ is the initial power spectrum. We assume $\P_\zeta$ is well
parameterized by $\P_\zeta = A_s (k/k_0)^{n_s-1}$ where $A_s$ gives the
normalization, $n_s$ is the scalar spectral index and $k_0$ is a
choice of normalization point.
Note that our number of degrees of freedom is actually less than
generic inflation, because although we have introduced $B$ we now no
longer have the amplitude and slope of the tensor component to consider.
The slope of the isocurvature perturbation is
predicted to be the same as the adiabatic perturbation and the tensors
are predicted to be negligible in the curvaton scenario \cite{Lyth02}.

The isocurvature modes have little effect on small scales,
but as
can be seen from \eq{SWsq} they can either raise {\em or lower} the
Sachs Wolfe plateau relative to the acoustic peaks depending on the
sign of $B$. This is in contrast to tensor perturbations which can
only raise the Sachs Wolfe plateau relative to the acoustic peaks.

\begin{figure}
\begin{center}
\epsfig{figure=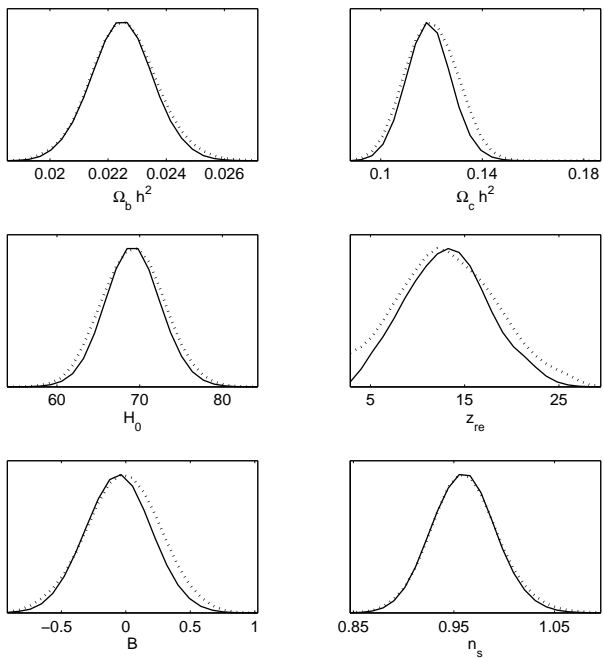,width=8cm}
\end{center}
\caption{\slabel{isocorr} 
Posterior marginalized probability
distributions (solid lines) of the cosmological parameters including
correlated matter isocurvature modes, using the data described in the
text.  $B$ is the ratio of the (effective) baryon isocurvature to
adiabatic perturbation amplitude in the primordial era,
$\Omegab h^2$ and $\Omegac h^2$ are the physical matter
densities in baryons and CDM, $H_0 \Hunit$ is the Hubble parameter
today, $z_{\text{re}}$ is the effective reionization redshift, and $n_s$ is the
spectral index. We assume a flat universe with cosmological constant.
Dotted lines show the mean likelihoods of the samples,
and agree well with the marginalized curves, indicating the full
distribution is fairly Gaussian and unskewed~\cite{Lewis02}.}
\end{figure}

\begin{figure}
\begin{center}
\epsfig{figure=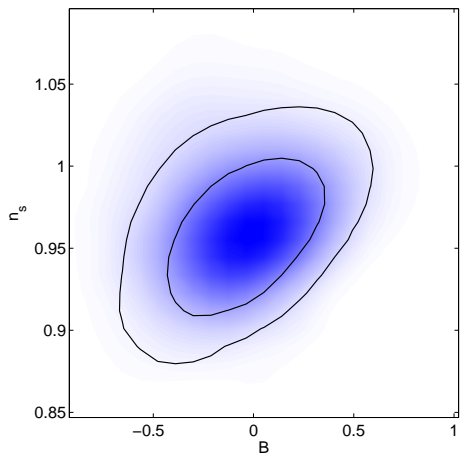,width=6.1cm}
\end{center}
\caption{
\label{Bns} 
Posterior distribution of $B = \Sbeff/\zeta$ in the
  primordial era, and the spectral index $n_s$. The plot is generated
  from a smoothed number density of Monte Carlo samples generated
  using the data described in the text. The contours enclose $68\%$
  and $95\%$ of the probability, and the shading is by the mean
  likelihood of the samples.
}
\end{figure}

Computing the full predictions numerically and assuming a flat prior
on $B$, \fig{isocorr} shows the posterior distribution for the various
cosmological parameters when the possibility of a totally correlated
mixture of matter isocurvature and adiabatic perturbations is
allowed. The posterior distribution of $B$ and $n_s$ is shown in \fig{Bns}, marginalized over the other parameters. On small scales the isocurvature modes have only a
small effect, so the main observational constraint comes from the
relative amplitudes of the large and small scale power. This is
partially degenerate with the spectral index as clearly demonstrated
in the figure. 
The relative large scale amplitude is also
affected by the reionization optical depth, and  although this is constrained
by WMAP's polarization measurements the experimental noise and cosmic variance still leave a
significant residual uncertainty.

We find the ratio of the mean likelihood allowing for isocurvature modes to
that for purely adiabatic models is about $0.7$ (for discussion of mean
likelihoods see Ref.~\cite{cosmomc}). Thus the isocurvature modes do not improve the already good
fit to the data of the standard purely adiabatic case. 
By the same token, the current data is still
consistent with a significant isocurvature contribution,
with the $95\%$ marginalized confidence interval $-0.53<B<0.43$. 
If new data
favored $B>0$ this would be largely degenerate with a tensor
contribution predicted by standard single field inflationary
scenarios, and would be hard to distinguish without good CMB
polarization data. 
Evidence for $B<0$ would be a smoking gun for an isocurvature mode,
though the large scale polarization data has large enough cosmic
variance that to distinguish it from an
adiabatic model with an unexpected initial power spectrum shape would
be difficult.

The $95\%$ confidence marginalized constraint on the spectral index $0.90<n_s<1.02$
translates into a constraint on the potential $V$ during inflation (in general a function
of the inflation field $\infl$ and the curvaton field $\curv$)~\cite{Lyth02}
\begin{equation}
-0.1 \alt 2 \left(\eta_{\curv\sigma} - \epsilon \right) \alt 0.02
\slabel{curvatonslope}
\end{equation}
where
\begin{equation}
\eta_{\curv\curv} \equiv \frac{\Mp^2}{V} \frac{\partial^2 V}{\partial
\curv^2} \quad\quad  \epsilon \equiv \frac{1}{2}\Mp^2\left(  \frac{1}{V} \frac{\partial
V}{\partial \infl}  \right)^2,
\end{equation}
$\Mp$ is the reduced Planck mass, and the quantities are evaluated at horizon crossing during
inflation. In standard inflationary models the potential
has to satisfy $V^{1/4} \sim 0.03 \epsilon^{1/4} \Mp$ to obtain the
correct fluctuation amplitude,  which is difficult without  using unnatural values of the
model parameters \cite{Dimopoulos01}. In the curvaton
scenario we assume the inflaton perturbations are negligible, and hence
the potential merely has to be much smaller than this number. These
conditions are therefore much easier
to satisfy with natural values for the model parameters in the
curvaton case \cite{Dimopoulos01}.  
In both
cases the 
inflaton component of the 
potential also has to provide more than about 60 e-folds of
inflation.

If  the CDM is created before the curvaton decays, and while the curvaton
still has negligible energy density,
its density is essentially unperturbed. After the curvature perturbation is
generated there is therefore a relative 
isocurvature 
perturbation, given by \cite{Lyth02}
\begin{equation}
\Sc \approx -3 \zeta.
\slabel{before}
\end{equation}
If the curvaton decays before its
energy density completely dominates, a CDM isocurvature
perturbation is produced~\cite{Lyth02}
\begin{equation}
\Sc \approx 3 \left( \frac{1 - r}{r} \right) \zeta,
\slabel{after}
\end{equation}
where $r$ measures the transfer function from
$\zeta_{\text{curvaton}}$ before curvaton decay to $\zeta$ after
decay, $
\zeta
 = r\zeta_{\text{curvaton}}$.  Ref.~\cite{Lyth02} find
the approximate result
$r \approx \rho_{\rm curvaton}/ \rho_{\rm total}$ where $\rho$ is the
energy density at curvaton decay,
to an accuracy of about 10\%~\cite{Malik02}.
 The same formulas, \eqs{before}{after}, apply for the
baryons with $\Sc$ replaced by $\Sb$.
If either the CDM or the baryon number was created after the curvaton
decayed then there would be no isocurvature perturbation in that
quantity \cite{Lyth02}. If both were created after the curvaton
decayed there would be no isocurvature modes.

There is no immediately compelling particle physics model for the
curvaton scenario~\cite{Postma02}, so
we consider nine basic scenarios depending on whether the CDM
and baryons are generated before, by, or after curvaton decay:
\begin{enumerate}
\item If both the CDM and baryon number is created after the curvaton
decay then there is no isocurvature perturbation:
\begin{equation}
B = 0.
\end{equation}
This scenario is consistent with the data
and indistinguishable from an inflation model with negligible tensor
component.

\item If the CDM is created before the curvaton decays and the baryon
number after the curvaton decays then from \eqs{before}{bcdm}
\begin{equation}
B = -3 \frac{\Rc}{\Rb}.
\end{equation}
This scenario is ruled out at high significance.
\item If the baryon number is created before the curvaton decays and
the CDM after the curvaton decays then from
\eq{before}
\begin{equation}
B = -3.
\end{equation}
This scenario is ruled out at high significance.
\item If the CDM is created by the curvaton decay and the baryon
number after the curvaton decays then from \eqs{after}{bcdm}
\begin{equation}
B = 3 \frac{\Rc}{\Rb} \left( \frac{1 - r}{r} \right).
\end{equation}
Solving for $r$ gives
\begin{equation}
r = \frac{1}{1 + (\Rb/\Rc)B/3}.
\end{equation}
\item If the baryon number is created by the curvaton decay and the
CDM after the curvaton decays then from \eq{after}
\begin{equation}
B = 3 \left( \frac{1 - r}{r} \right).
\end{equation}
Solving for $r$ gives
\begin{equation}
r = \frac{1}{1 + B/3}.
\end{equation}
\item 
If the CDM and baryons are both created before the curvaton decays
then from \eqs{bcdm}{before}
\begin{equation}
B = -\frac{3}{\Rb}.
\end{equation}
This scenario is ruled out at high significance.
\item
If CDM and baryons were both created by the curvaton then from
\eqs{bcdm}{after}
\begin{equation}
B = \frac{3}{\Rb}\frac{1-r}{r}.
\end{equation}
Solving for $r$ gives
\begin{equation}
r = \frac{1}{1+\Rb B/3}.
\slabel{r1}
\end{equation}
\item If the CDM is created before the curvaton decay and the baryons
are created by the curvaton decay then from \eqss{bcdm}{before}{after}
\begin{equation}
B = \frac{3(\Rb-r)}{r\Rb}.
\end{equation}
Solving for $r$ gives
\begin{equation}
r = \frac{\Rb}{1+\Rb B/3}.
\slabel{r2}
\end{equation}
\item
If the CDM is created by the curvaton and the baryons are created
before the curvaton decays we have 
from \eqss{bcdm}{before}{after}
\begin{equation}
B = \frac{3(\Rc-r)}{r\Rb}.
\end{equation}
Solving for $r$ gives
\begin{equation}
r = \frac{\Rc}{1+\Rb B/3}.
\slabel{r3}
\end{equation}
\end{enumerate}
For the cases that are not immediately ruled out we obtain a
constraint on $r$.
The posterior probability distribution for this quantity can easily be
constructed from the Monte Carlo samples, and a plot of it is shown in
\fig{r} for the various cases. The peaks at $r=1$ are when there
are no isocurvature modes. The curves which peak at $r\sim R_b$ and
$r\sim R_c$ are when compensating baryon and CDM isocurvature modes
are created before and by curvaton decay, giving a total effective
isocurvature perturbation close to zero.

\begin{figure}
\begin{center}
\epsfig{figure=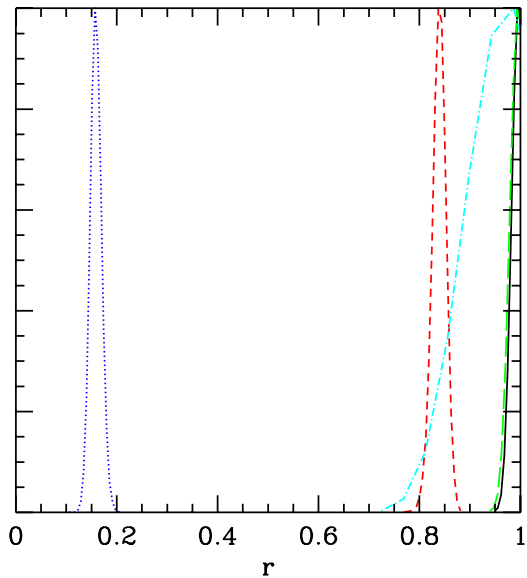,width=8cm}
\end{center}
\caption{\slabel{r} Plots of the un-normalized posterior probability distribution
for $r \approx \rho_{\rm curvaton}/ \rho_{\rm total}$ 
when the curvaton decays.
The distributions are for the numbered scenarios described in the text:
(4) CDM created by curvaton decay and baryon number after curvaton decay
(\web{green }long dashes),
(5) baryon number created by curvaton decay and CDM after curvaton
decay (\web{cyan }dash-dot line),
(7) both CDM and baryon number created by curvaton decay (\web{black }solid line),
(8) CDM created before curvaton decay and
baryon number by curvaton decay (\web{blue }dotted line),
(9) baryon number created before curvaton decay and CDM by curvaton
decay (\web{red }short dashed line).
}
\end{figure}
\begin{figure}
\begin{center}
\includegraphics[width=8cm]{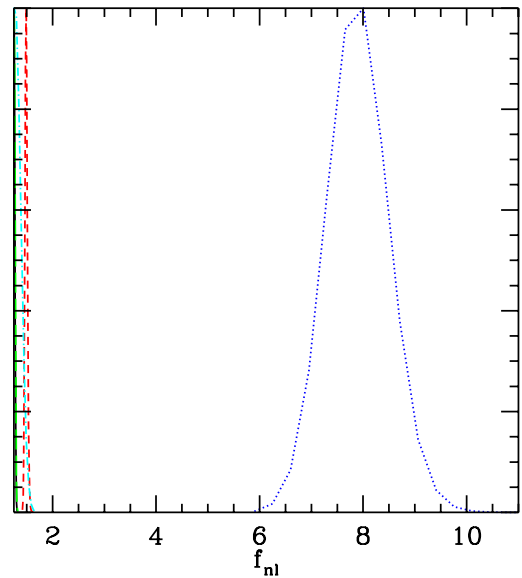}
\end{center}
\caption{\label{fig:fnl} 
Plots of the un-normalized posterior probability distribution
for the amount of non-Gaussianity, $f_{nl}$. The line styles are the
same as in \fig{r}.}
\end{figure}

The amount of non-Gaussianity in the CMB is dependent on $r$ 
with the conventional governing parameter
\cite{Lyth02}
\begin{equation}
f_{nl} \approx \frac{5}{4r}
\end{equation}
assuming $f_{nl} \gg 1$.
Using this equation we can convert the likelihood plots for $r$ into
those for $f_{nl}$ as is shown in \fig{fig:fnl}.
The values near $f_{nl}\sim 1$ should not be taken too seriously as
there will be additional second order non-Gaussian contributions from
fields other than the curvaton.
The current one year WMAP data has $f_{nl}<134$ $(95\%)$ and is predicted to
reach $f_{nl} < 80$ $(95\%)$ with the four year WMAP data
\cite{Komatsu03}.
So if WMAP eventually
detects non-Gaussianity it will rule out all the models considered
here.
The Planck satellite
 is predicted to  ultimately be able to detect $f_{nl} \gtrsim
5$  \cite{Komatsu01}.
 If this is realized Planck will be able to distinguish between the case where
the CDM is created before curvaton decay and the baryon number by
curvaton decay and the other possibilities.

\section{Conclusions}
\label{conclusions}

The curvaton model provides a simple scenario that can give rise to
correlated adiabatic and isocurvature modes of similar size.  
The current data 
do not favor a large isocurvature contribution,
but a significant amplitude is still allowed.

We point out that the CDM and baryon isocurvature modes differ only by
the addition of an observationally null mode in which the two
perturbations compensate. The
CDM isocurvature mode can therefore be treated
as a scaled baryon isocurvature mode. 
A simple analytical approximation for the effect of mixtures of
large scale
isocurvature and adiabatic perturbations on the CMB temperature
anisotropy was given.
Numerically, we found that the data was consistent at the  two sigma level with the presence of an effective correlated  baryon isocurvature perturbation of about 50\% the magnitude 
of the adiabatic perturbation.
The individual baryon and CDM isocurvature modes can be even larger
if they compensate each other.
Models in which
either the baryon number or CDM was created 
 before the curvaton 
dominated the energy density
are
ruled out unless counter-balanced by the other species being created
by the curvaton decay.
The levels of non-Gaussianity expected for the various scenarios
were evaluated and in the case of the CDM being created before the
curvaton decayed and the baryon number by the curvaton decay, could be
high enough to detectable by the Planck satellite.

\begin{acknowledgments}
AL thanks Anthony Challinor for some very useful notes and the Kavli
Institute of Theoretical Physics where part of this work was done. 
CG thanks David Lyth and David Wands for helpful discussions.  
The Beowulf computer used for this analysis was funded by the Canada 
Foundation for Innovation and the Ontario Innovation Trust.
This research was supported in part by the National Science Foundation
 under Grant No. PHY94-07194 and by PPARC (UK).

\end{acknowledgments}

\appendix

\section{Covariant perturbation equations}
\label{cov_eqs}
The covariant approach to cosmological perturbation theory gives a set of gauge
invariant equations in which all the terms are
covariant and have a physical interpretation~\cite{Ellis83,Challinor99}. 
The quantities can be calculated in any 
frame (labelled by a
4-velocity $u_a$) and the equations remain the same. Individual
quantities measuring a particular perturbation do in general depend on
what frame $u_a$ is used to calculate them, so when talking about (for
example) a
density perturbation it is important to make clear what frame one is
referring to.

The spatial gradient of the 3-Ricci scalar ${}^{(3)}\clr$ vanishes in a
homogeneous universe, and  $\eta_a = \half S D_a
{}^{(3)}\clr$ is a natural covariant measure of the scalar curvature
perturbation in some frame with  4-velocity
$u_a$. Here $S$ is the scale factor and $D_a$ is the spatial covariant
derivative orthogonal to $u_a$ (we 
use the signature where $u_a u^a=1$).
Other covariant quantities useful for studying perturbations are
defined in \cite{Challinor99,Lewis02}, along with derivations of the
equations of
General Relativity that relate them. Here we only consider scalar
modes at linear order in a spatially flat universe\footnote{The
  equations given here generalize trivially to a non-flat universe by
  the substitution $\eta \rightarrow \eta/(1-3K/k^2)$.}, and perform a
harmonic expansion as described in \cite{Challinor99}, leaving the
$k$-dependence of scalar quantities implicit. 
For example we describe
the curvature perturbation by the scalar harmonic coefficient $\eta$.

Frame invariant quantities can be constructed from combinations of covariant
quantities that depend on the choice of frame $u_a$. 
These often have an interpretation in terms of the value of a
particular quantity in some specified frame. 
In particular
\begin{equation}
\Phi \equiv \half\eta + \frac{\H\sigma}{k},
\end{equation}
where $\sigma$ is the scalar shear and $\H = 3 S \nabla^a u_a$ is the conformal Hubble parameter, is proportional to the curvature
perturbation 
in the zero shear frame (the Newtonian gauge). The acceleration $A$ in the
zero shear frame 
\begin{equation}
\Psi \equiv -A + (\sigma' + \H \sigma)/k
\end{equation} 
defines a second frame invariant quantity, which is related to $\Phi$  by
\begin{equation}
\Phi + \Psi = -\frac{\kappa S^2 \Pi}{k^2}
\end{equation}
where $\Pi$ is the anisotropic stress. The Weyl tensor is the part of
the Riemann tensor which is not determined by the local stress-energy,
and defines a 
frame independent
scalar potential\footnote{The $\Phi$ of
  Ref.~\cite{Challinor99} has a different sign convention where $\Phi \equiv -\phi$.} $\phi$ \cite{Challinor99} which is related to the above
via
\begin{equation}
\phi = \half(\Psi - \Phi).
\end{equation}
We define a frame invariant curvature perturbation 
\begin{equation}
\zeta \equiv \frac{\eta}{2} - \frac{\H\clx}{\rho'},
\end{equation}
proportional to the curvature perturbation in the uniform density
frame. Here $\clx = \sum_i \clx_i = \sum_i \rho_i\Delta_i$ is the total
density perturbation. This is 
related to the comoving curvature perturbation
\begin{equation}
\chi \equiv -\frac{\eta}{2} + \frac{\H q}{k(\rho+p)} = -\left[\Phi +\frac{2}{3}\frac{\H^{-1}\Phi' - \Psi}{1+w}\right].
\label{chi_eq}
\end{equation}
by $\chi = -\zeta - \H \bar{\clx}/\rho'$, where $\bar{\clx}$ is the
comoving density perturbation and $q= \sum_i (\rho_i+p_i)v_i$ is
the total heat flux and $v_i$ are the velocities. The Poisson equation relates the density and
potential via $k^2\Phi = \half \kappa S^2 \bar{\clx}$. It follows
that for adiabatic modes where $\chi$ is non-zero initially $\chi
\approx -\zeta$ on large scales.

A local scale factor $S$ can be defined (up to an initial value) by
integrating the local expansion rate $\H$,
and the quantity $h_a \equiv D_a S = S D_a S/S$ (scalar harmonic coefficient $h$) provides a measure of the
perturbation to local volume elements. The derivative $h'$ with
respect to conformal time $\tau$ is
unambiguously defined, and describes the rate of change of local
volume element perturbations. In the frame in which $h'$ is zero fractional perturbations in number
densities of conserved species remain constant if there are no matter flows.
The evolution of the curvature perturbation is given by
\begin{equation}
\eta' = 2h' -\frac{2}{3}k\sigma 
\label{etadot}
\end{equation}
so on large scales the $h'=0$ frame coincides with the $\eta'=0$
frame. Thus $\eta$ is conserved on large scales in the frame in which
number density perturbation fractions are constant~\cite{Lyth02}. This
result is purely a result of linear torsionless spacetime geometry.

The time evolution of the local scale factor perturbation sources growth of
density perturbations of uninteracting conserved species via the
energy conservation equation
\begin{equation}
\clx_i' + 3\H(\clx_i + \delta p_i) + k (\rho_i+p_i)v_i = -3 h'(\rho_i+p_i),
\label{energy_cons}
\end{equation}
where  $\delta p_i$ is the  pressure perturbation.
The $h'=0$ frame therefore
coincides with the $\clx_i=0$ frame on large scales if $\delta p_i=0$
in the $\clx_i=0$ frame. For a particular species one can define the
curvature perturbation in the frame in which its density is unperturbed
\begin{equation}
\zeta_i \equiv  \frac{\eta}{2} -  \frac{\H\clx_i}{\rho_i'},
\end{equation}
where in the absence of energy transfer $\rho_i' = -3\H(\rho_i+p_i)$.
The evolution equation that follows from \eqs{energy_cons}{etadot}
is~\cite{Wands00}
\begin{equation}
\zeta_i' = -\frac{\H}{\rho_i+p_i}\left[ \delta p_i -
  \frac{p'}{\rho'}\clx_i\right] - \frac{k V_i}{3}, 
\label{zetadot}
\end{equation}
where $V_i \equiv v_i+\sigma$ is the Newtonian gauge velocity. If there is an
equation of state $p_i=p_i(\rho_i)$ the first term on the right hand
side is zero, and the $\zeta_i$ are therefore constant on large scales where $k\tau \ll 1$. If the equation
of state parameter $w_i\equiv p_i/\rho_i$ is constant this implies that 
the fractional density perturbations in the unperturbed curvature
frame evolve as
\begin{equation}
\Drho_i' = -k (1+w_i) V_i,
\end{equation}
and hence the $\Drho_i$ are also conserved on large scales. The
curvature perturbation in the frame in which the total energy is
unperturbed is given from the $\zeta_i$ by \eq{zetasum}.
In the frame in which the acceleration $A=0$ (and hence $u_a$
coincides with the CDM velocity) $\eta_s = -\eta/2$, $h_s' = 6h'$
where $\eta_s$ and $h_s'$ are the synchronous gauge quantities
(e.g. see \cite{Ma95}).

\section{Isocurvature initial conditions}
\label{series}
In the early radiation dominated era there are in general five
regular solutions to the perturbation
equations~\cite{Bucher99}, assuming there is only one distinct species
of cold dark matter. If there are several species of dark matter the
additional modes are unobservable without measuring the distinct dark
matter species directly. Performing a series expansion in conformal time
$\tau$, the Friedman equation gives
\begin{equation}
S = \frac{\Omega_m H_0^2}{\omega^2}\left( \omega\tau +
\frac{1}{4}\omega^2\tau^2 + \clo( K\omega \tau^3) \right)
\end{equation}
where $\omega\equiv \Omega_m\H_0/\sqrt{\Omega_\gamma+\Omega_\nu}$
with $\H_0$ the Hubble parameter today and $\Omega_i$ the density today in units of the critical density. 
At lowest order in the tight coupling
expansion, assuming the baryons and dark matter have negligible
pressure, the CDM isocurvature mode at early times is
\ba
\Drho_c &=& 1- \frac{1}{72}\frac{\Rc(4 \Rv-15) \omega k^2
  \tau^3}{2 \Rv +15}\\
\Drho_\gamma &=& \Drho_\nu = \frac{4}{3} \Drho_b=  \frac{5}{6}\frac{\Rc \omega k^2
  \tau^3}{2(\Rv+15)}\\
V_c  &=& \frac{1}{24}\frac{\Rc(4 \Rv-15) \omega k \tau^2}{2 \Rv+15}\\
V_\gamma &=& V_\nu= V_b =  -\frac{15}{8} \frac{\Rc \omega k \tau^2}{2 \Rv+15}
\\
\Psi &=& \frac{1}{8} \frac{\Rc (4 \Rv-15) \omega\tau}{2 \Rv+15} \\
\Phi &=& \frac{1}{8} \frac{\Rc (4 \Rv+15) \omega\tau}{2 \Rv+15}
\ea
where equalities apply at the given order in $\tau$. 
The baryon isocurvature mode is given by subtracting the
observationally null mode
$\Drho_c = - \Rb\Drho_b/\Rc$ from the above solution. Series solutions
for the adiabatic and isocurvature modes to any order are easily
computed using computer algebra packages, for a Maple derivation of the
 solutions in the zero acceleration frame see
 \url{http://camb.info/theory.html}. The above solution was calculated
 by constructing the frame invariant quantities above from the
 quantities in the zero acceleration frame. 

The $\Drho_i$ are constant to order $(k\tau)^2$. However the
lowest order terms in the velocities are of order $(\omega \tau)(k
\tau)$, demonstrating explicitly that the assumption that $V_i =
\text{const}\times k\tau + \clo((k\tau)^2)$ in \cite{Langlois00} is
incorrect for isocurvature modes.
\vfill


\begin{thebibliography}{67}
\expandafter\ifx\csname natexlab\endcsname\relax\def\natexlab#1{#1}\fi
\expandafter\ifx\csname bibnamefont\endcsname\relax
  \def\bibnamefont#1{#1}\fi
\expandafter\ifx\csname bibfnamefont\endcsname\relax
  \def\bibfnamefont#1{#1}\fi
\expandafter\ifx\csname citenamefont\endcsname\relax
  \def\citenamefont#1{#1}\fi
\expandafter\ifx\csname url\endcsname\relax
  \def\url#1{\texttt{#1}}\fi
\expandafter\ifx\csname urlprefix\endcsname\relax\def\urlprefix{URL }\fi
\providecommand{\bibinfo}[2]{#2}
\providecommand{\eprint}[2][]{\url{#2}}

\bibitem[{\citenamefont{{G. Hinshaw et al.}}(2003)}]{Hinshaw03}
\bibinfo{author}{\bibnamefont{{G. Hinshaw et al.}}} (\bibinfo{year}{2003}),
  \eprint{astro-ph/0302217}.

\bibitem[{\citenamefont{{H. V. Peiris et al.}}(2003)}]{Peiris03}
\bibinfo{author}{\bibnamefont{{H. V. Peiris et al.}}} (\bibinfo{year}{2003}),
  \eprint{astro-ph/0302225}.

\bibitem[{\citenamefont{Liddle and Lyth}(2000)}]{LL}
\bibinfo{author}{\bibfnamefont{A.}~\bibnamefont{Liddle}} \bibnamefont{and}
  \bibinfo{author}{\bibfnamefont{D.}~\bibnamefont{Lyth}},
  \emph{\bibinfo{title}{Cosmological Inflation And Large-Scale Structure}}
  (\bibinfo{publisher}{Cambridge University Press}, \bibinfo{year}{2000}).

\bibitem[{\citenamefont{Kofman and Linde}(1987)}]{Kofman87}
\bibinfo{author}{\bibfnamefont{L.~A.} \bibnamefont{Kofman}} \bibnamefont{and}
  \bibinfo{author}{\bibfnamefont{A.~D.} \bibnamefont{Linde}},
  \bibinfo{journal}{Nucl. Phys. B} \textbf{\bibinfo{volume}{282}},
  \bibinfo{pages}{555} (\bibinfo{year}{1987}).

\bibitem[{\citenamefont{Polarski and Starobinsky}(1994)}]{Polarski94}
\bibinfo{author}{\bibfnamefont{D.}~\bibnamefont{Polarski}} \bibnamefont{and}
  \bibinfo{author}{\bibfnamefont{A.~A.} \bibnamefont{Starobinsky}},
  \bibinfo{journal}{Phys. Rev. D} \textbf{\bibinfo{volume}{50}},
  \bibinfo{pages}{6123} (\bibinfo{year}{1994}),
  \eprint[http://arXiv.org/abs]{astro-ph/9404061}.

\bibitem[{\citenamefont{Garcia-Bellido and Wands}(1996)}]{Garcia-Bellido96}
\bibinfo{author}{\bibfnamefont{J.}~\bibnamefont{Garcia-Bellido}}
  \bibnamefont{and} \bibinfo{author}{\bibfnamefont{D.}~\bibnamefont{Wands}},
  \bibinfo{journal}{Phys. Rev. D} \textbf{\bibinfo{volume}{53}},
  \bibinfo{pages}{5437} (\bibinfo{year}{1996}),
  \eprint[http://arXiv.org/abs]{astro-ph/9511029}.

\bibitem[{\citenamefont{Linde and Mukhanov}(1997)}]{Linde97}
\bibinfo{author}{\bibfnamefont{A.}~\bibnamefont{Linde}} \bibnamefont{and}
  \bibinfo{author}{\bibfnamefont{V.}~\bibnamefont{Mukhanov}},
  \bibinfo{journal}{Phys. Rev. D} \textbf{\bibinfo{volume}{56}},
  \bibinfo{pages}{535} (\bibinfo{year}{1997}),
  \eprint[http://arXiv.org/abs]{astro-ph/9610219}.

\bibitem[{\citenamefont{Sasaki and Tanaka}(1998)}]{Sasaki98}
\bibinfo{author}{\bibfnamefont{M.}~\bibnamefont{Sasaki}} \bibnamefont{and}
  \bibinfo{author}{\bibfnamefont{T.}~\bibnamefont{Tanaka}},
  \bibinfo{journal}{Prog. Theor. Phys.} \textbf{\bibinfo{volume}{99}},
  \bibinfo{pages}{763} (\bibinfo{year}{1998}),
  \eprint[http://arXiv.org/abs]{gr-qc/9801017}.

\bibitem[{\citenamefont{Langlois}(1999)}]{Langlois99}
\bibinfo{author}{\bibfnamefont{D.}~\bibnamefont{Langlois}},
  \bibinfo{journal}{Phys. Rev. D} \textbf{\bibinfo{volume}{59}},
  \bibinfo{pages}{123512} (\bibinfo{year}{1999}), \eprint{astro-ph/9906080}.

\bibitem[{\citenamefont{Gordon et~al.}(2001)\citenamefont{Gordon, Wands,
  Bassett, and Maartens}}]{Gordon00}
\bibinfo{author}{\bibfnamefont{C.}~\bibnamefont{Gordon}},
  \bibinfo{author}{\bibfnamefont{D.}~\bibnamefont{Wands}},
  \bibinfo{author}{\bibfnamefont{B.~A.} \bibnamefont{Bassett}},
  \bibnamefont{and} \bibinfo{author}{\bibfnamefont{R.}~\bibnamefont{Maartens}},
  \bibinfo{journal}{Phys. Rev. D} \textbf{\bibinfo{volume}{63}},
  \bibinfo{pages}{023506} (\bibinfo{year}{2001}),
  \eprint[http://arXiv.org/abs]{astro-ph/0009131}.

\bibitem[{\citenamefont{Hwang and Noh}(2000)}]{Hwang00}
\bibinfo{author}{\bibfnamefont{J.-C.} \bibnamefont{Hwang}} \bibnamefont{and}
  \bibinfo{author}{\bibfnamefont{H.}~\bibnamefont{Noh}},
  \bibinfo{journal}{Phys. Lett. B} \textbf{\bibinfo{volume}{495}},
  \bibinfo{pages}{277} (\bibinfo{year}{2000}),
  \eprint[http://arXiv.org/abs]{astro-ph/0009268}.

\bibitem[{\citenamefont{Wands et~al.}(2002)\citenamefont{Wands, Bartolo,
  Matarrese, and Riotto}}]{Wands02}
\bibinfo{author}{\bibfnamefont{D.}~\bibnamefont{Wands}},
  \bibinfo{author}{\bibfnamefont{N.}~\bibnamefont{Bartolo}},
  \bibinfo{author}{\bibfnamefont{S.}~\bibnamefont{Matarrese}},
  \bibnamefont{and} \bibinfo{author}{\bibfnamefont{A.}~\bibnamefont{Riotto}},
  \bibinfo{journal}{Phys. Rev. D} \textbf{\bibinfo{volume}{66}},
  \bibinfo{pages}{043520} (\bibinfo{year}{2002}),
  \eprint[http://arXiv.org/abs]{astro-ph/0205253}.

\bibitem[{\citenamefont{Tsujikawa et~al.}(2002)\citenamefont{Tsujikawa,
  Parkinson, and Bassett}}]{Tsujikawa02}
\bibinfo{author}{\bibfnamefont{S.}~\bibnamefont{Tsujikawa}},
  \bibinfo{author}{\bibfnamefont{D.}~\bibnamefont{Parkinson}},
  \bibnamefont{and} \bibinfo{author}{\bibfnamefont{B.~A.}
  \bibnamefont{Bassett}} (\bibinfo{year}{2002}),
  \eprint[http://arXiv.org/abs]{astro-ph/0210322}.

\bibitem[{\citenamefont{Starobinsky et~al.}(2001)\citenamefont{Starobinsky,
  Tsujikawa, and Yokoyama}}]{Starobinsky01}
\bibinfo{author}{\bibfnamefont{A.~A.} \bibnamefont{Starobinsky}},
  \bibinfo{author}{\bibfnamefont{S.}~\bibnamefont{Tsujikawa}},
  \bibnamefont{and} \bibinfo{author}{\bibfnamefont{J.}~\bibnamefont{Yokoyama}},
  \bibinfo{journal}{Nucl. Phys.} \textbf{\bibinfo{volume}{B610}},
  \bibinfo{pages}{383} (\bibinfo{year}{2001}), \eprint{astro-ph/0107555}.

\bibitem[{\citenamefont{Bartolo et~al.}(2001)\citenamefont{Bartolo, Matarrese,
  and Riotto}}]{Bartolo01}
\bibinfo{author}{\bibfnamefont{N.}~\bibnamefont{Bartolo}},
  \bibinfo{author}{\bibfnamefont{S.}~\bibnamefont{Matarrese}},
  \bibnamefont{and} \bibinfo{author}{\bibfnamefont{A.}~\bibnamefont{Riotto}},
  \bibinfo{journal}{Phys. Rev.} \textbf{\bibinfo{volume}{D64}},
  \bibinfo{pages}{123504} (\bibinfo{year}{2001}), \eprint{astro-ph/0107502}.

\bibitem[{\citenamefont{Groot~Nibbelink and van Tent}(2002)}]{GrootNibbelink01}
\bibinfo{author}{\bibfnamefont{S.}~\bibnamefont{Groot~Nibbelink}}
  \bibnamefont{and} \bibinfo{author}{\bibfnamefont{B.~J.~W.} \bibnamefont{van
  Tent}}, \bibinfo{journal}{Class. Quant. Grav.} \textbf{\bibinfo{volume}{19}},
  \bibinfo{pages}{613} (\bibinfo{year}{2002}), \eprint{hep-ph/0107272}.

\bibitem[{\citenamefont{Mollerach}(1990)}]{Mollerach90}
\bibinfo{author}{\bibfnamefont{S.}~\bibnamefont{Mollerach}},
  \bibinfo{journal}{Phys. Rev. D} \textbf{\bibinfo{volume}{42}},
  \bibinfo{pages}{313} (\bibinfo{year}{1990}).

\bibitem[{\citenamefont{Lyth and Wands}(2002)}]{Lyth01}
\bibinfo{author}{\bibfnamefont{D.}~\bibnamefont{Lyth}} \bibnamefont{and}
  \bibinfo{author}{\bibfnamefont{D.}~\bibnamefont{Wands}},
  \bibinfo{journal}{Phys. Lett. B} \textbf{\bibinfo{volume}{524}},
  \bibinfo{pages}{5} (\bibinfo{year}{2002}), \eprint{hep-ph/0110002}.

\bibitem[{\citenamefont{Lyth et~al.}(2003)\citenamefont{Lyth, Ungarelli, and
  D.Wands}}]{Lyth02}
\bibinfo{author}{\bibfnamefont{D.~H.}~\bibnamefont{Lyth}},
  \bibinfo{author}{\bibfnamefont{C.}~\bibnamefont{Ungarelli}},
  \bibnamefont{and} \bibinfo{author}{\bibnamefont{D.Wands}},
  \bibinfo{journal}{Phys. Rev. D} \textbf{\bibinfo{volume}{67}},
  \bibinfo{pages}{023503} (\bibinfo{year}{2003}), \eprint{astro-ph/0208055}.

\bibitem[{\citenamefont{Moroi and Takahashi}(2001)}]{Moroi01}
\bibinfo{author}{\bibfnamefont{T.}~\bibnamefont{Moroi}} \bibnamefont{and}
  \bibinfo{author}{\bibfnamefont{T.}~\bibnamefont{Takahashi}},
  \bibinfo{journal}{Phys. Lett. B} \textbf{\bibinfo{volume}{522}},
  \bibinfo{pages}{215} (\bibinfo{year}{2001}), \eprint{hep-ph/0110096}.

\bibitem[{\citenamefont{Moroi and Takahashi}(2002)}]{Moroi02}
\bibinfo{author}{\bibfnamefont{T.}~\bibnamefont{Moroi}} \bibnamefont{and}
  \bibinfo{author}{\bibfnamefont{T.}~\bibnamefont{Takahashi}},
  \bibinfo{journal}{Phys. Rev. D} \textbf{\bibinfo{volume}{66}},
  \bibinfo{pages}{063501} (\bibinfo{year}{2002}), \eprint{hep-ph/0206026}.

\bibitem[{\citenamefont{Moroi and Murayama}(2003)}]{Murayam02}
\bibinfo{author}{\bibfnamefont{T.}~\bibnamefont{Moroi}} \bibnamefont{and}
  \bibinfo{author}{\bibfnamefont{H.}~\bibnamefont{Murayama}},
  \bibinfo{journal}{Phys. Lett. B} \textbf{\bibinfo{volume}{553}},
  \bibinfo{pages}{126} (\bibinfo{year}{2003}), \eprint{hep-ph/0211019}.

\bibitem[{\citenamefont{Dimopoulos and Lyth}(2002)}]{Dimopoulos01}
\bibinfo{author}{\bibfnamefont{K.}~\bibnamefont{Dimopoulos}} \bibnamefont{and}
  \bibinfo{author}{\bibfnamefont{D.}~\bibnamefont{Lyth}}
  (\bibinfo{year}{2002}), \eprint[http://arXiv.org/abs]{hep-ph/0209180}.

\bibitem[{\citenamefont{Bastero-Gil et~al.}(2002)\citenamefont{Bastero-Gil,
  Di~Clemente, and King}}]{Bastero-Gil02}
\bibinfo{author}{\bibfnamefont{M.}~\bibnamefont{Bastero-Gil}},
  \bibinfo{author}{\bibfnamefont{V.}~\bibnamefont{Di~Clemente}},
  \bibnamefont{and} \bibinfo{author}{\bibfnamefont{S.~F.} \bibnamefont{King}}
  (\bibinfo{year}{2002}), \eprint[http://arXiv.org/abs]{hep-ph/0211011}.

\bibitem[{\citenamefont{Postma}(2003)}]{Postma02}
\bibinfo{author}{\bibfnamefont{M.}~\bibnamefont{Postma}},
  \bibinfo{journal}{Phys. Rev. D} \textbf{\bibinfo{volume}{67}},
  \bibinfo{pages}{063518} (\bibinfo{year}{2003}), \eprint{hep-ph/0212005}.

\bibitem[{\citenamefont{Enqvist et~al.}(2003)\citenamefont{Enqvist, Kasuya, and
  Mazumdar}}]{Enqvist02}
\bibinfo{author}{\bibfnamefont{K.}~\bibnamefont{Enqvist}},
  \bibinfo{author}{\bibfnamefont{S.}~\bibnamefont{Kasuya}}, \bibnamefont{and}
  \bibinfo{author}{\bibfnamefont{A.}~\bibnamefont{Mazumdar}},
  \bibinfo{journal}{Phys. Rev. Lett} \textbf{\bibinfo{volume}{90}},
  \bibinfo{pages}{091302} (\bibinfo{year}{2003}), \eprint{hep-ph/0211147}.

\bibitem[{\citenamefont{Dimopoulos
  et~al.}(2003{\natexlab{a}})\citenamefont{Dimopoulos, Lazarides, Lyth, and
  Ruiz~de Austri}}]{Dimopoulos03a}
\bibinfo{author}{\bibfnamefont{K.}~\bibnamefont{Dimopoulos}},
  \bibinfo{author}{\bibfnamefont{G.}~\bibnamefont{Lazarides}},
  \bibinfo{author}{\bibfnamefont{D.}~\bibnamefont{Lyth}}, \bibnamefont{and}
  \bibinfo{author}{\bibfnamefont{R.}~\bibnamefont{Ruiz~de Austri}}
  (\bibinfo{year}{2003}{\natexlab{a}}), \eprint{hep-ph/0303154}.

\bibitem[{\citenamefont{Dimopoulos
  et~al.}(2003{\natexlab{b}})\citenamefont{Dimopoulos, Lyth, Notari, and
  Riotto}}]{Dimopoulos03b}
\bibinfo{author}{\bibfnamefont{K.}~\bibnamefont{Dimopoulos}},
  \bibinfo{author}{\bibfnamefont{D.~H.} \bibnamefont{Lyth}},
  \bibinfo{author}{\bibfnamefont{A.}~\bibnamefont{Notari}}, \bibnamefont{and}
  \bibinfo{author}{\bibfnamefont{A.}~\bibnamefont{Riotto}}
  (\bibinfo{year}{2003}{\natexlab{b}}), \eprint{hep-ph/0304050}.

\bibitem[{\citenamefont{Enqvist and Sloth}(2002)}]{Enqvist01}
\bibinfo{author}{\bibfnamefont{K.}~\bibnamefont{Enqvist}} \bibnamefont{and}
  \bibinfo{author}{\bibfnamefont{M.~S.} \bibnamefont{Sloth}},
  \bibinfo{journal}{Nucl. Phys.} \textbf{\bibinfo{volume}{B626}},
  \bibinfo{pages}{395} (\bibinfo{year}{2002}), \eprint{hep-ph/0109214}.

\bibitem[{\citenamefont{Sloth}(2003)}]{Sloth02}
\bibinfo{author}{\bibfnamefont{M.~S.} \bibnamefont{Sloth}},
  \bibinfo{journal}{Nucl. Phys. B} \textbf{\bibinfo{volume}{656}},
  \bibinfo{pages}{239} (\bibinfo{year}{2003}), \eprint{hep-ph/0208241}.

\bibitem[{\citenamefont{Bozza et~al.}(2002)\citenamefont{Bozza, Gasperini,
  Giovannini, and Veneziano}}]{Bozza02}
\bibinfo{author}{\bibfnamefont{V.}~\bibnamefont{Bozza}},
  \bibinfo{author}{\bibfnamefont{M.}~\bibnamefont{Gasperini}},
  \bibinfo{author}{\bibfnamefont{M.}~\bibnamefont{Giovannini}},
  \bibnamefont{and}
  \bibinfo{author}{\bibfnamefont{G.}~\bibnamefont{Veneziano}},
  \bibinfo{journal}{Phys. Lett.} \textbf{\bibinfo{volume}{B543}},
  \bibinfo{pages}{14} (\bibinfo{year}{2002}), \eprint{hep-ph/0206131}.

\bibitem[{\citenamefont{Copeland et~al.}(1997)\citenamefont{Copeland, Easther,
  and Wands}}]{Copeland97}
\bibinfo{author}{\bibfnamefont{E.~J.} \bibnamefont{Copeland}},
  \bibinfo{author}{\bibfnamefont{R.}~\bibnamefont{Easther}}, \bibnamefont{and}
  \bibinfo{author}{\bibfnamefont{D.}~\bibnamefont{Wands}},
  \bibinfo{journal}{Phys. Rev.} \textbf{\bibinfo{volume}{D56}},
  \bibinfo{pages}{874} (\bibinfo{year}{1997}), \eprint{hep-th/9701082}.

\bibitem[{\citenamefont{Lidsey et~al.}(2000)\citenamefont{Lidsey, Wands, and
  Copeland}}]{Lidsey00}
\bibinfo{author}{\bibfnamefont{J.~E.} \bibnamefont{Lidsey}},
  \bibinfo{author}{\bibfnamefont{D.}~\bibnamefont{Wands}}, \bibnamefont{and}
  \bibinfo{author}{\bibfnamefont{E.~J.} \bibnamefont{Copeland}},
  \bibinfo{journal}{Phys. Rept.} \textbf{\bibinfo{volume}{337}},
  \bibinfo{pages}{343} (\bibinfo{year}{2000}), \eprint{hep-th/9909061}.

\bibitem[{\citenamefont{Efstathiou and Bond}(1986)}]{Efstathiou86}
\bibinfo{author}{\bibfnamefont{G.}~\bibnamefont{Efstathiou}} \bibnamefont{and}
  \bibinfo{author}{\bibfnamefont{J.~R.} \bibnamefont{Bond}},
  \bibinfo{journal}{{Mon. Not. R. Astron. Soc.}}
  \textbf{\bibinfo{volume}{218}}, \bibinfo{pages}{103} (\bibinfo{year}{1986}).

\bibitem[{\citenamefont{{R. Stompor et al.}}(1996)}]{Stompor96}
\bibinfo{author}{\bibnamefont{{R. Stompor et al.}}},
  \bibinfo{journal}{Astrophys. J.} \textbf{\bibinfo{volume}{463}}
  (\bibinfo{year}{1996}).

\bibitem[{\citenamefont{Pierpaoli et~al.}(1999)\citenamefont{Pierpaoli,
  Garcia-Bellido, and Borgani}}]{Pierpaoli99}
\bibinfo{author}{\bibfnamefont{E.}~\bibnamefont{Pierpaoli}},
  \bibinfo{author}{\bibfnamefont{J.}~\bibnamefont{Garcia-Bellido}},
  \bibnamefont{and} \bibinfo{author}{\bibfnamefont{S.}~\bibnamefont{Borgani}},
  \bibinfo{journal}{JHEP} \textbf{\bibinfo{volume}{10}}, \bibinfo{pages}{015}
  (\bibinfo{year}{1999}), \eprint[http://arXiv.org/abs]{hep-ph/9909420}.

\bibitem[{\citenamefont{Kawasaki and Takahashi}(2001)}]{Kawasaki2001}
\bibinfo{author}{\bibfnamefont{M.}~\bibnamefont{Kawasaki}} \bibnamefont{and}
  \bibinfo{author}{\bibfnamefont{F.}~\bibnamefont{Takahashi}},
  \bibinfo{journal}{Phys. Lett.} \textbf{\bibinfo{volume}{B516}},
  \bibinfo{pages}{388} (\bibinfo{year}{2001}),
  \eprint[http://arXiv.org/abs]{hep-ph/0105134}.

\bibitem[{\citenamefont{Enqvist et~al.}(2000)\citenamefont{Enqvist,
  Kurki-Suonio, and Valiviita}}]{Enqvist2000}
\bibinfo{author}{\bibfnamefont{K.}~\bibnamefont{Enqvist}},
  \bibinfo{author}{\bibfnamefont{H.}~\bibnamefont{Kurki-Suonio}},
  \bibnamefont{and}
  \bibinfo{author}{\bibfnamefont{J.}~\bibnamefont{Valiviita}},
  \bibinfo{journal}{Phys. Rev.} \textbf{\bibinfo{volume}{D62}},
  \bibinfo{pages}{103003} (\bibinfo{year}{2000}),
  \eprint[http://arXiv.org/abs]{astro-ph/0006429}.

\bibitem[{\citenamefont{Langlois and Riazuelo}(2000)}]{Langlois00}
\bibinfo{author}{\bibfnamefont{D.}~\bibnamefont{Langlois}} \bibnamefont{and}
  \bibinfo{author}{\bibfnamefont{A.}~\bibnamefont{Riazuelo}},
  \bibinfo{journal}{Phys. Rev. D} \textbf{\bibinfo{volume}{62}},
  \bibinfo{pages}{043504} (\bibinfo{year}{2000}), \eprint{astro-ph/9912497}.

\bibitem[{\citenamefont{Bucher et~al.}(2001{\natexlab{a}})\citenamefont{Bucher,
  Moodley, and Turok}}]{Bucher99}
\bibinfo{author}{\bibfnamefont{M.}~\bibnamefont{Bucher}},
  \bibinfo{author}{\bibfnamefont{K.}~\bibnamefont{Moodley}}, \bibnamefont{and}
  \bibinfo{author}{\bibfnamefont{N.}~\bibnamefont{Turok}},
  \bibinfo{journal}{Phys. Rev. D} \textbf{\bibinfo{volume}{62}},
  \bibinfo{pages}{083508} (\bibinfo{year}{2001}{\natexlab{a}}),
  \eprint{astro-ph/9904231}.

\bibitem[{\citenamefont{Trotta et~al.}(2001)\citenamefont{Trotta, Riazuelo, and
  Durrer}}]{Trotta01}
\bibinfo{author}{\bibfnamefont{R.}~\bibnamefont{Trotta}},
  \bibinfo{author}{\bibfnamefont{A.}~\bibnamefont{Riazuelo}}, \bibnamefont{and}
  \bibinfo{author}{\bibfnamefont{R.}~\bibnamefont{Durrer}},
  \bibinfo{journal}{Phys. Rev. Lett.} \textbf{\bibinfo{volume}{87}},
  \bibinfo{pages}{231301} (\bibinfo{year}{2001}),
  \eprint[http://arXiv.org/abs]{astro-ph/0104017}.

\bibitem[{\citenamefont{Trotta et~al.}(2003)\citenamefont{Trotta, Riazuelo, and
  Durrer}}]{Trotta02}
\bibinfo{author}{\bibfnamefont{R.}~\bibnamefont{Trotta}},
  \bibinfo{author}{\bibfnamefont{A.}~\bibnamefont{Riazuelo}}, \bibnamefont{and}
  \bibinfo{author}{\bibfnamefont{R.}~\bibnamefont{Durrer}},
  \bibinfo{journal}{Phys. Rev. D} \textbf{\bibinfo{volume}{67}},
  \bibinfo{pages}{063520} (\bibinfo{year}{2003}), \eprint{astro-ph/0211600}.

\bibitem[{\citenamefont{Bucher et~al.}(2001{\natexlab{b}})\citenamefont{Bucher,
  Moodley, and Turok}}]{Bucher01}
\bibinfo{author}{\bibfnamefont{M.}~\bibnamefont{Bucher}},
  \bibinfo{author}{\bibfnamefont{K.}~\bibnamefont{Moodley}}, \bibnamefont{and}
  \bibinfo{author}{\bibfnamefont{N.}~\bibnamefont{Turok}},
  \bibinfo{journal}{Phys. Rev. Lett.} \textbf{\bibinfo{volume}{87}},
  \bibinfo{pages}{191301} (\bibinfo{year}{2001}{\natexlab{b}}),
  \eprint{astro-ph/0012141}.

\bibitem[{\citenamefont{Amendola et~al.}(2002)\citenamefont{Amendola, Gordon,
  Wands, and Sasaki}}]{Amendola01}
\bibinfo{author}{\bibfnamefont{L.}~\bibnamefont{Amendola}},
  \bibinfo{author}{\bibfnamefont{C.}~\bibnamefont{Gordon}},
  \bibinfo{author}{\bibfnamefont{D.}~\bibnamefont{Wands}}, \bibnamefont{and}
  \bibinfo{author}{\bibfnamefont{M.}~\bibnamefont{Sasaki}},
  \bibinfo{journal}{Phys. Rev. Lett.} \textbf{\bibinfo{volume}{88}},
  \bibinfo{pages}{211302} (\bibinfo{year}{2002}),
  \eprint[http://arXiv.org/abs]{astro-ph/0107089}.

\bibitem[{\citenamefont{Bartolo and Liddle}(2002)}]{Bartolo02}
\bibinfo{author}{\bibfnamefont{N.}~\bibnamefont{Bartolo}} \bibnamefont{and}
  \bibinfo{author}{\bibfnamefont{A.~R.} \bibnamefont{Liddle}},
  \bibinfo{journal}{Phys. Rev.} \textbf{\bibinfo{volume}{D65}},
  \bibinfo{pages}{121301} (\bibinfo{year}{2002}), \eprint{astro-ph/0203076}.

\bibitem[{\citenamefont{{L. Verde et al.}}(2003)}]{Verde03}
\bibinfo{author}{\bibnamefont{{L. Verde et al.}}} (\bibinfo{year}{2003}),
  \eprint{astro-ph/0302218}.

\bibitem[{\citenamefont{{A. Kogut et al.}}(2003)}]{Kogut03}
\bibinfo{author}{\bibnamefont{{A. Kogut et al.}}} (\bibinfo{year}{2003}),
  \eprint{astro-ph/0302213}.

\bibitem[{\citenamefont{{C.L. Kuo et al.}}(2002)}]{Kuo02}
\bibinfo{author}{\bibnamefont{{C.L. Kuo et al.}}} (\bibinfo{year}{2002}),
  \eprint{astro-ph/0212289}.

\bibitem[{\citenamefont{{W. Percival et al.}}(2002)}]{Percival02}
\bibinfo{author}{\bibnamefont{{W. Percival et al.}}}, \bibinfo{journal}{MNRAS}
  \textbf{\bibinfo{volume}{337}}, \bibinfo{pages}{1068} (\bibinfo{year}{2002}),
  \eprint{astro-ph/0206256}.

\bibitem[{\citenamefont{{W. L. Freedman et al.}}(2001)}]{Freedman01}
\bibinfo{author}{\bibnamefont{{W. L. Freedman et al.}}},
  \bibinfo{journal}{{\apj}} \textbf{\bibinfo{volume}{553}}, \bibinfo{pages}{47}
  (\bibinfo{year}{2001}), \eprint{astro-ph/0012376}.

\bibitem[{\citenamefont{Burles et~al.}(2001)\citenamefont{Burles, Nollett, and
  Turner}}]{Burles01}
\bibinfo{author}{\bibfnamefont{S.}~\bibnamefont{Burles}},
  \bibinfo{author}{\bibfnamefont{K.~M.} \bibnamefont{Nollett}},
  \bibnamefont{and} \bibinfo{author}{\bibfnamefont{M.~S.}
  \bibnamefont{Turner}}, \bibinfo{journal}{Astrophys. J.}
  \textbf{\bibinfo{volume}{552}}, \bibinfo{pages}{L1} (\bibinfo{year}{2001}),
  \eprint{astro-ph/0010171}.

\bibitem[{\citenamefont{Lewis and Bridle}(2002)}]{cosmomc}
\bibinfo{author}{\bibfnamefont{A.}~\bibnamefont{Lewis}} \bibnamefont{and}
  \bibinfo{author}{\bibfnamefont{S.}~\bibnamefont{Bridle}},
  \bibinfo{journal}{Phys. Rev. D} \textbf{\bibinfo{volume}{66}},
  \bibinfo{pages}{103511} (\bibinfo{year}{2002}), \eprint{astro-ph/0205436}.

\bibitem[{\citenamefont{Bardeen}(1980)}]{Bardeen80}
\bibinfo{author}{\bibfnamefont{J.~M.} \bibnamefont{Bardeen}},
  \bibinfo{journal}{Phys. Rev. D} \textbf{\bibinfo{volume}{22}},
  \bibinfo{pages}{1882} (\bibinfo{year}{1980}).

\bibitem[{\citenamefont{Bardeen et~al.}(1983)\citenamefont{Bardeen, Steinhardt,
  and Turner}}]{BST}
\bibinfo{author}{\bibfnamefont{J.~M.} \bibnamefont{Bardeen}},
  \bibinfo{author}{\bibfnamefont{P.~J.} \bibnamefont{Steinhardt}},
  \bibnamefont{and} \bibinfo{author}{\bibfnamefont{M.~S.}
  \bibnamefont{Turner}}, \bibinfo{journal}{Phys. Rev. D}
  \textbf{\bibinfo{volume}{28}}, \bibinfo{pages}{679} (\bibinfo{year}{1983}).

\bibitem[{\citenamefont{Lyth}(1985)}]{Lyth85}
\bibinfo{author}{\bibfnamefont{D.~H.} \bibnamefont{Lyth}},
  \bibinfo{journal}{Phys. Rev. D} \textbf{\bibinfo{volume}{31}},
  \bibinfo{pages}{1792} (\bibinfo{year}{1985}).

\bibitem[{\citenamefont{Kodama and Sasaki}(1984)}]{Kodama84}
\bibinfo{author}{\bibfnamefont{H.}~\bibnamefont{Kodama}} \bibnamefont{and}
  \bibinfo{author}{\bibfnamefont{M.}~\bibnamefont{Sasaki}},
  \bibinfo{journal}{Prog. Theor. Phys. Suppl.} \textbf{\bibinfo{volume}{78}},
  \bibinfo{pages}{1} (\bibinfo{year}{1984}).

\bibitem[{\citenamefont{Mukhanov et~al.}(1992)\citenamefont{Mukhanov, Feldman,
  and Brandenberger}}]{Mukhanov92}
\bibinfo{author}{\bibfnamefont{V.}~\bibnamefont{Mukhanov}},
  \bibinfo{author}{\bibfnamefont{H.}~\bibnamefont{Feldman}}, \bibnamefont{and}
  \bibinfo{author}{\bibfnamefont{R.}~\bibnamefont{Brandenberger}},
  \bibinfo{journal}{Phys. Rep.} \textbf{\bibinfo{volume}{215}},
  \bibinfo{pages}{203} (\bibinfo{year}{1992}).

\bibitem[{\citenamefont{Wands et~al.}(2000)\citenamefont{Wands, Malik, Lyth,
  and Liddle}}]{Wands00}
\bibinfo{author}{\bibfnamefont{D.}~\bibnamefont{Wands}},
  \bibinfo{author}{\bibfnamefont{K.~A.} \bibnamefont{Malik}},
  \bibinfo{author}{\bibfnamefont{D.~H.} \bibnamefont{Lyth}}, \bibnamefont{and}
  \bibinfo{author}{\bibfnamefont{A.~R.} \bibnamefont{Liddle}},
  \bibinfo{journal}{Phys. Rev. D} \textbf{\bibinfo{volume}{62}},
  \bibinfo{pages}{043527} (\bibinfo{year}{2000}), \eprint{astro-ph/0003278}.

\bibitem[{\citenamefont{Malik et~al.}(2003)\citenamefont{Malik, Wands, and
  Ungarelli}}]{Malik02}
\bibinfo{author}{\bibfnamefont{K.~H.}~\bibnamefont{Malik}},
  \bibinfo{author}{\bibfnamefont{D.}~\bibnamefont{Wands}}, \bibnamefont{and}
  \bibinfo{author}{\bibfnamefont{C.}~\bibnamefont{Ungarelli}},
  \bibinfo{journal}{Phys. Rev. D} \textbf{\bibinfo{volume}{67}},
  \bibinfo{pages}{063516} (\bibinfo{year}{2003}), \eprint{astro-ph/0211602}.

\bibitem[{\citenamefont{Seljak and Zaldarriaga}(1996)}]{Seljak96}
\bibinfo{author}{\bibfnamefont{U.}~\bibnamefont{Seljak}} \bibnamefont{and}
  \bibinfo{author}{\bibfnamefont{M.}~\bibnamefont{Zaldarriaga}},
  \bibinfo{journal}{Astrophys. J.} \textbf{\bibinfo{volume}{469}},
  \bibinfo{pages}{437} (\bibinfo{year}{1996}), \eprint{astro-ph/9603033}.

\bibitem[{\citenamefont{Lewis et~al.}(2000)\citenamefont{Lewis, Challinor, and
  Lasenby}}]{Lewis99}
\bibinfo{author}{\bibfnamefont{A.}~\bibnamefont{Lewis}},
  \bibinfo{author}{\bibfnamefont{A.}~\bibnamefont{Challinor}},
  \bibnamefont{and} \bibinfo{author}{\bibfnamefont{A.}~\bibnamefont{Lasenby}},
  \bibinfo{journal}{Astrophys. J.} \textbf{\bibinfo{volume}{538}},
  \bibinfo{pages}{473} (\bibinfo{year}{2000}),
  \bibinfo{note}{{\url{http://camb.info}}}, \eprint{astro-ph/9911177}.

\bibitem[{\citenamefont{Lewis and Challinor}(2002)}]{Lewis02}
\bibinfo{author}{\bibfnamefont{A.}~\bibnamefont{Lewis}} \bibnamefont{and}
  \bibinfo{author}{\bibfnamefont{A.}~\bibnamefont{Challinor}},
  \bibinfo{journal}{Phys. Rev. D} \textbf{\bibinfo{volume}{66}},
  \bibinfo{pages}{023531} (\bibinfo{year}{2002}), \eprint{astro-ph/0203507}.

\bibitem[{\citenamefont{{E. Komatsu et. al}}(2003)}]{Komatsu03}
\bibinfo{author}{\bibnamefont{{E. Komatsu et. al}}} (\bibinfo{year}{2003}),
  \eprint{astro-ph/0302223}.

\bibitem[{\citenamefont{Komatsu and Spergel}(2001)}]{Komatsu01}
\bibinfo{author}{\bibfnamefont{E.}~\bibnamefont{Komatsu}} \bibnamefont{and}
  \bibinfo{author}{\bibfnamefont{D.~N.} \bibnamefont{Spergel}},
  \bibinfo{journal}{Phys. Rev. D} \textbf{\bibinfo{volume}{63}},
  \bibinfo{pages}{063002} (\bibinfo{year}{2001}), \eprint{astro-ph/0005036}.

\bibitem[{\citenamefont{Ellis et~al.}(1983)\citenamefont{Ellis, Matravers, and
  Treciokas}}]{Ellis83}
\bibinfo{author}{\bibfnamefont{G.~F.~R.} \bibnamefont{Ellis}},
  \bibinfo{author}{\bibfnamefont{D.~R.} \bibnamefont{Matravers}},
  \bibnamefont{and}
  \bibinfo{author}{\bibfnamefont{R.}~\bibnamefont{Treciokas}},
  \bibinfo{journal}{Ann. Phys.} \textbf{\bibinfo{volume}{150}},
  \bibinfo{pages}{455} (\bibinfo{year}{1983}).

\bibitem[{\citenamefont{Challinor and Lasenby}(1999)}]{Challinor99}
\bibinfo{author}{\bibfnamefont{A.}~\bibnamefont{Challinor}} \bibnamefont{and}
  \bibinfo{author}{\bibfnamefont{A.}~\bibnamefont{Lasenby}},
  \bibinfo{journal}{Astrophys. J.} \textbf{\bibinfo{volume}{513}},
  \bibinfo{pages}{1} (\bibinfo{year}{1999}), \eprint{astro-ph/9804301}.

\bibitem[{\citenamefont{Ma and Bertschinger}(1995)}]{Ma95}
\bibinfo{author}{\bibfnamefont{C.-P.} \bibnamefont{Ma}} \bibnamefont{and}
  \bibinfo{author}{\bibfnamefont{E.}~\bibnamefont{Bertschinger}},
  \bibinfo{journal}{\apj} \textbf{\bibinfo{volume}{455}}, \bibinfo{pages}{7}
  (\bibinfo{year}{1995}), \eprint{astro-ph/9506072}.

\end{thebibliography}

\end{document}